\documentclass[preprint,12pt]{elsarticle}

\usepackage{amssymb}

\usepackage{siunitx}

\journal{Optics Communications}

\graphicspath{ {./} }

\begin{document}

\begin{frontmatter}



\title{Using polarization sensitive SMLM to infer the interaction strength of dye-plasmonic nanosphere systems}


\author[1]{T. Novák\corref{cor1}}\ead{novaktibor@titan.physx.u-szeged.hu}
\author[1]{P. Bíró}
\author[2]{Gy. Ferenc}
\author[3]{D. Ungor}
\author[1]{E. Czvik}
\author[4]{Á. Deák}
\author[4]{L. Janovák}
\author[1]{M. Erdélyi\corref{cor1}}
\ead{erdelyi.miklos@szte.hu}

\cortext[cor1]{Corresponding author}

\affiliation[1]{organization={Department of Optics and Quantum Electronics, University of Szeged},
            addressline={Dóm tér 9},
            city={Szeged},
            postcode={6720},
            country={Hungary}}

\affiliation[2]{organization={Institute of Plant Biology, Biological Research Centre, Hungarian Research Network},
            addressline={Temesvári krt. 62},
            city={Szeged},
            postcode={6720},
            country={Hungary}}

\affiliation[3]{organization={MTA-SZTE Lendület "Momentum" Noble Metal Nanostructures Research Group, University of Szeged},
            addressline={Rerrich B. sqr. 1},
            city={Szeged},
            postcode={6720},
            country={Hungary}}

\affiliation[4]{organization={Interdisciplinary Excellence Center, Department of Physical Chemistry and Materials Science, University of Szeged},
            addressline={Rerrich B. sqr. 1},
            city={Szeged},
            postcode={6720},
            country={Hungary}}

\begin{abstract}

Single molecule microscopy has proven to be an effective tool to characterize the fluorophore-plasmonic structure interaction. However, as specific information is hidden in the emission, sophisticated evaluation is required. Here we investigated the emission polarization of rotationally mobile fluorophores near plasmonic Au/Ag alloy nanospheres both theoretically and experimentally. Our work surpasses the results of the previous studies in that it considers the rotational mobility of the fluorophores attached to metallic nanoparticles. Through theoretical modeling and elaborate numerical calculations we determined the expected measurable fluorescence polarization, and via DNA-PAINT single molecule measurements we validated the theoretical predictions. Our findings suggest that through measuring the polarization state of a sequence of single-molecule events, it is possible to infer the interaction strength of the dye-nanosphere system. A precise description requires elaborate calculations, however, a simplified model is able to predict the tendency of the polarization state with the change of experimental parameters.

\end{abstract}




\end{frontmatter}


\section{Introduction}
\label{sec:introduction}


Plasmonic nanoparticles (NPs) possess many unique size- and composition-dependent properties which originate from their strong interaction with light. This can manifest in scattering and absorption with cross sections exceeding the geometrical size or in greatly enhanced locally confined electric fields. Moreover, the interaction is highly wavelength dependent, and is sensitive to the surrounding medium\cite{olson_optical_2015}. These properties facilitate various applications in the field of sensing\cite{mejia-salazar_plasmonic_2018,taylor_single-molecule_2017}, biomedicine\cite{bai_goldsilver_2017}, SERS\cite{liu_shell_2012,chen_picomolar_2013}, catalysis\cite{zeng_comparison_2010,zou_imaging_2018}, photothermal applications\cite{chen_gold_2010,hu_efficient_2009}, and they are promising basic materials for drug delivery systems\cite{yamashita_controlled-release_2011,simoncelli2015thermoplasmonic}. The diversity of the applications requires the synthesis of nanoparticles with specific properties. Seed mediated growth\cite{bai_goldsilver_2017}, which is one of the most popular chemical methods, enables the synthesis of nanoparticles having varied compositions and shapes.

However, the properties of the produced nanoparticles often sensitively depend on the synthesis parameters, which cannot be sufficiently controlled. Consequently, it is indispensable to examine and characterize the synthetized particles. For this purpose, the most widely used procedure is geometrical characterization by scanning (SEM) and transmission electron microscopy (TEM). However, these techniques do not serve direct information about the plasmonic properties of the nanoparticles, therefore some form of light-matter interaction is required.

By measuring the excitation or scattered spectrum, valuable information can be obtained about the average plasmonic properties of the noble metal NPs. Cathode luminescence\cite{vesseur2007direct}, electron energy loss spectroscopy and near-field scanning optical microscopy\cite{betzig_near_1986} allow for probing the local field around the particles. The latter even enables the examination of individual particles, but the probe itself perturbs the field. The investigation of fluorophore–nanoparticle interaction requires a different approach, and fluorescence microscopy seems to be a good candidate. The earliest such methods measured only the ensemble fluorescence properties\cite{geddes_metal_2002} (e.g. the lifetime, brightness, or spectrum). For this reason, preference was given to single-molecule localization microscopy (SMLM) techniques capable of recording the signals of individual fluorophores and hence providing much richer information.

The SMLM methods, which generally create images with sub-diffraction resolution by fitting the point spread functions (PSF) of individual fluorescent molecules, have become well established to this day\cite{mockl_super-resolution_2020,schermelleh_super-resolution_2019}. One of these methods is the DNA-PAINT technique\cite{jungmann_single-molecule_2010,simoncelli2018imaging}, which uses short, diffusing, labeled ssDNA imaging strands as fluorescent probes, while the complementary docking sequences are bound to the target structure. By adjusting the hybridization kinetics of the imaging and docking strands, appropriate fluorescence blinking can be achieved for SMLM. To minimize the fluorescent background, total internal reflection (TIR) illumination is commonly used as the labels remain fluorescent while diffusing.

Although the SMLM techniques are most frequently used for measuring the molecular organization of biological samples, they have also been found useful in other fields, e.g. for the investigation of plasmonic nanoparticles\cite{willets_super-resolution_2017,chattopadhyay_super-resolution_2021}. On the one hand, the geometry of the individual nanoparticles can be mapped, however, it has been shown that the real particle sizes can only be retrieved under special conditions because the apparent size strongly depends on the plasmonic interaction\cite{blythe_tripletstatemediated_2014,fu2017super,goldwyn_mislocalization_2018,taylor_all-optical_2018}. On the other hand, measuring quantities other than molecule coordinates (e.g. brightness, polarization, lifetime), the fluorescence modification of individual fluorophores can be directly quantified\cite{chattopadhyay_super-resolution_2021,mack_decoupling_2017,steuwe_visualizing_2015,toth_mapping_2020,zuo_rotation_2019,blanquer2020relocating}.

The fluorescent radiation of dyes can be described by dipole emission, basically polarized radiation, which is seldom experienced in SMLM experiments as the sample is labeled with fluorophores through rotary and flexible linkers\cite{backer_enhanced_2016}. Therefore, the orientation of the dyes averages out during the exposure time. However, there are samples where more direct labeling can be applied with fixed or constrained fluorophores\cite{ries_superresolution_2013}. In the case of these samples, a polarization-sensitive optical element must be inserted in the emission path to separate the polarization components or to modify the PSF\cite{valades_cruz_quantitative_2016,sinko_polarization_2017}, by which the orientation of the fluorophores can be determined, and structural information can be obtained about the sample.

In this work we performed elaborate calculations on the polarization state of fluorophore emission attached to Au-Ag alloyed spherical nanoparticles. We showed that the emission of the fluorophore-spherical nanopraticle system is highly polarized even when the fluorophore has complete rotational freedom. Utilizing the developed framework, we investigated the wavelength, nanoparticle size and position dependence of the measurable polarization degree. Furthermore, we performed DNA-PAINT experiments using Atto647N and AF488 dyes on Au-Ag alloyed nanoparticles. The measurements show the highly polarized nature of the single molecule emissions and confirm our theoretical predictions. Our results point out that the strength of the dye-nanoparticle interaction can be inferred from the measurable polarization degree (MPD) distribution of the single molecule blinking events.

\section{Methods}
\label{sec:methods}

\subsection{Theoretical background}
\label{sec:methods_theoretical_backgroud}

To predict the polarization state of the emission, we performed elaborate numerical calculations taking into account the size of the nanosphere, the dye’s rotational freedom and its relative position on the nanosphere. The Python code used for the calculations is available at our Gitlab repository\cite{noauthor_mie_nodate}.

\subsubsection{Geometry}
\label{sec:methods_geometry}

In the model, the metal nanosphere was embedded in water medium (Figure \ref{fig:methods_theoretical}A). The effect of the refractive index mismatch of the glass coverslip and of the DNA strands was neglected. The refractive index of the alloyed nanosphere was estimated by weighted average of the electric susceptibilities\cite{lee_gold_2006} of gold and silver\cite{johnson_optical_1972}, based on the composition ratio. The composition ratio was chosen to be 30\% gold and 70\% silver. The dyes, acting as dipole sources, were placed on a cropped spherical surface \qty{18}{\nm} away from the nanoparticle surface (Figure \ref{fig:methods_theoretical}A, B). The emitted radiation was collected in the far-field region with a collection angle of \ang{78.6} corresponding to the numerical aperture of the imaging system.

\subsubsection{Mie theory}
\label{sec:methods_mie_theory}

The calculation of the fluorophore–Au/Ag alloy nanosphere interaction is based on the Mie theory. The theory can describe both the interaction of a plane wave (excitation) and a dipole source (emission) with a spherical object. Regarding the excitation, the p-polarized TIR illumination was approximated with a plane wave polarized in the "z" direction. For dipole sources near a spherical object, the scattered and total electric fields (Supplementary Eq. 1 and 2), the excited state decay rate enhancements (Supplementary Eq. 3) and the quantum efficiency (Supplementary Eq. 5) can be calculated using analytical formulas\cite{kerker_surface_1980,ruppin_decay_1982,chew_transition_1987,mertens_plasmon-enhanced_2007}. The wavelengths in the calculations were chosen according to the measurement conditions, i.e. \qty{488}{nm} and \qty{647}{\nm} wavelengths were used for the plane wave excitation, and \qty{520}{\nm} and \qty{670}{\nm} wavelengths were used for the dipole sources corresponding to the AF488 and Atto647N dyes, respectively.

\subsubsection{Brightness of emitters}
\label{sec:methods_brightness}

Although the Mie theory can describe the emission power of a dipole emitter, it does not directly serves the emission power of real fluorophores. In the fluorophores’ linear regime, their brightness is limited by the excitation process and is unaffected by the relaxation rate. During the measurements, the excitation laser fields were moderate, even with local intensity enhancement around the nanoparticle (maximum $\sim 15 \times$), and they remained in the dye’s linear regime. Apart from the excitation strength, we considered the orientation dependent quantum efficiency (Figure \ref{fig:methods_theoretical}C) as well as the orientation and position dependent collection efficiency.

\subsubsection{Orientation averaging}
\label{sec:methods_orientation_averaging}

The rotational mobility of the dyes attached to the ssDNA strands was taken into account by averaging the dipole orientations of the fluorophore. We assumed that the dyes rotate freely, as the AF488 does not stick to the end of the DNA strands and the  rotational correlation time of the Atto647 is still orders of magnitude shorter than the camera exposure\cite{vandenberk_evaluation_2018}. This means that during one camera exposure ($\sim\qty{100}{\ms}$) all possible dipole orientations need to be considered with equal weight. It is further assumed the dyes rotate unconstrained and averaging can be performed to the whole \qty{4}{\pi} solid angle. The details of the averaging process are implemented as the limiting case of slow rotation, where the characteristic time relations of the fluorescence, rotation and camera exposure are:
$$
\tau_{fluorescence} \ll \tau_{rotation} \ll \tau_{frame}\mathrm{.}
$$

In this limit, the dye’s fluorescence lifetime is significantly shorter than the rotational correlation time, and the dipole orientations during excitation and emission are taken as the same. The basis of this assumption is twofold. Firstly, according to the control measurements (see Section \ref{sec:results_experiments}), without the plasmonic interaction the two temporal constants are roughly the same for the slower rotating Atto647N dye. Secondly, the interaction with the particle significantly decreases the fluorescence lifetime (due to the radiative rate enhancement and the decrease of the quantum efficiency) for most geometry and dye orientation (Figure \ref{fig:methods_theoretical}C). Although this assumption is not valid when the dipole is close to parallel to the nanoparticle surface, the emission quantum efficiency is also highly reduced in this case, so these orientations contribute less to the average signal. In the fast rotation limiting case the degree of polarization is typically higher by 10--40\%  (Supplementary Fig. S1).

All evaluations and numerical calculations in the study were performed using the slow rotation limiting case. The collected emission signals of the two polarization channels in this limiting case can be expressed as:
$$
{signal} ^ {x , y} (\vec{r}) \sim {mean}_{\vec{d}} \left[\frac{QE}{\sigma_{s}} {\left| {E}_{local} (\vec{r}) \vec{d} \right|} ^ {2} \int_{ap} {{\left|E_{farfield}^{x , y} (\vec{r} , \vec {d}) \right|} ^ {2}} \,df \right] \mathrm{,}
$$
where $E_{local}$ is the local excitation field of the incident plane wave, $E_{farfield}$ is the emitted far field of an ideal dipole moment, $QE$ is the quantum yield and $\sigma_s$ is the scattering rate of the emission for the given orientation. For the fast rotation case, see Supplementary Eq. 6 and 7.

\subsubsection{Degree of polarization and Monte Carlo simulations}
\label{sec:methods_monte_carlo}

\begin{figure}[h]
\centering
\includegraphics[width=\linewidth]{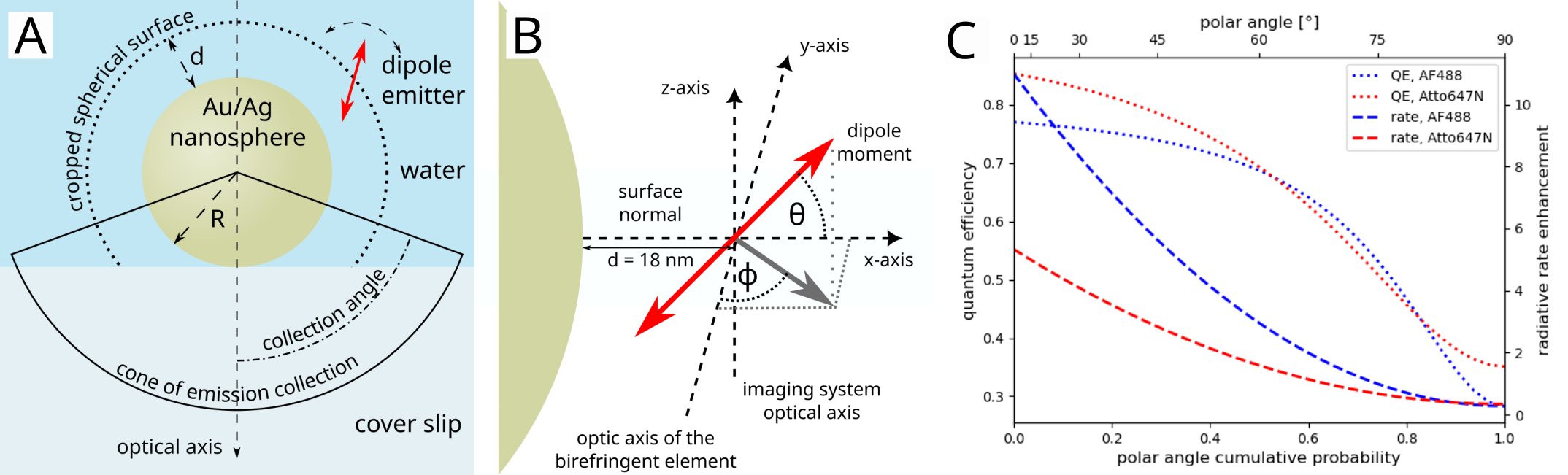}
\caption{A: The geometric model used in the study. The dyes are attached to the surface of spherical nanoparticles with a radius $R$ at a fixed $d$ distance. B: The convention of the coordinate system. It depicts the case when the dye is placed on the "x" axis, thus the surface normal aligns with this axis. The $\theta$ and the $\phi$ are the polar and azimuth angles, respectively. C: The emission quantum yields (dotted lines) and radiative rate enhancements (dashed lines) as a function of the dipole polar angle in case of a \qty{80}{\nm} diameter particle. The cumulative probability gives the probability that a polar angle of the dipole moment is less than a given $\theta$ value.}
\label{fig:methods_theoretical}
\end{figure}

So far, modeling has been confined to a fluorophore in a single position, corresponding to a single blinking event. However, the MPD highly depends on the relative alignment of the polarizing element’s axes (here aligned along the "x" and "y" axes) and the place on the nanosphere where the dye binds, which cannot be controlled experimentally. Consequently, instead of a single polarization degree value we measure a distribution. The stochastic nature of the binding events was taken into account by Monte Carlo simulations during which the dye positions on the cropped sphere are chosen randomly. In each position, the orientations are averaged and the MPD is determined. From the many ($\sim10,000$) simulated blinking events histograms are created that can directly be juxtaposed with the histograms of the measurements. The MPD values of a blinking event are calculated as follows:
$$
MPD(\vec{r}) = \frac{
{signal}^{x}(\vec{r}) - {signal}^{y} (\vec{r})
}{
{signal}^{x} (\vec{r}) + {signal}^{y}(\vec{r})
}
$$
The MPD can take values between $-1$ and $1$. In contrast, the degree of polarization provides information on the polarization level of the emission, regardless of the alignment of the polarizing element:
$$
P(\vec{r}) = \frac{
{signal}_{polarized}(\vec{r})
}{
{signal}_{polarized}(\vec{r}) + {signal}_{unpolarized}(\vec{r})
}
$$
The higher propensity of surface-parallel dipole mode generation (related to the radiative rate enhancement, Figure \ref{fig:methods_theoretical}C) compared to that of the surface perpendicular dipole mode makes the emission more polarized, while the rotation of the dye and the presence of higher order modes depolarizes the emission. In the calculations, we exploited the rotational symmetry of the nanoparticles to equate the degree of polarization to the absolute MPD value of dyes placed in the "x-z" or the "y-z" planes.

\subsection{Experiments}
\label{sec:experiments}

\subsubsection{Synthesis of the Au/Ag alloyed colloidal nanoparticles}
\label{sec:experiments_nanoparticle_synthesis}

The synthesis of size-controlled Au/Ag alloy NPs was performed by using Au seed solution based on the protocol of Rioux\cite{rioux_seeded_2015}. Firstly, the gold seeds were prepared using a standard Turkevich method: \qty{300}{\uL} \qty{30}{mM} $\mathrm{HAuCl_4}$ was added to \qty{28}{\mL} MQ-water (\qty{18}{M \Omega \cdot{} cm}) in a closed Duran\textsuperscript{\textregistered{}} flask and heated to \qty{100}{\degreeCelsius}. While stirring the solution, we rapidly added \qty{200}{\uL} \qty{170}{mM} trisodium citrate solution. After a few minutes, the solution turned red. After the appearance of the red color, the solution was stirred and kept at \qty{100}{\degreeCelsius} for \qty{30}{\minute}. Finally, MQ-water was added to the solution to adjust the total volume to \qty{30}{\mL}. As a first step towards the preparation of alloy NPs, \qty{5}{\mL} seed solution was diluted with \qty{93.5}{\mL} MQ-water. The solution was stirred (\qty{550}{rpm}) and heated to \qty{90}{\degreeCelsius}. After heating, the adequate amount of $\mathrm{HAuCl_4}$ (\qty{30}{mM}) and $\mathrm{AgNO_3}$ (\qty{30}{mM}) were added to obtain the desired composition. For example, to produce Au/Ag:30/70 alloy NPs we put \qty{135}{\uL} \qty{30}{mM} $\mathrm{HAuCl_4}$ and \qty{315}{\uL} $\mathrm{AgNO_3}$ into the mixture. To reach the final volume of \qty{100}{\mL}, \qty{900}{\uL} \qty{170}{mM} trisodium citrate was added to the flask. The solution was left stirring for \qty{60}{\minute} at \qty{90}{\degreeCelsius}. In each successive growing step the alloy colloids from the previous step were used as seeds. Between two successive growing steps the colloids were purified by centrifugation at \qty{13,000}{rpm} for \qty{20}{\minute}.

\subsubsection{Synthesis and purification of thiol-modified and fluorescently labeled DNA oligonucleotides}
\label{sec:experiments_oligonucleotid_synthesis}

Phosphoramidites, synthesis reagents and solvents used for DNA oligonucleotide synthesis were purchased from Sigma-Aldrich Kft. (Budapest, Hungary), Molar Chemicals Kft. (Budapest, Hungary) and LGC Genomics Ltd. (Teddington, UK). AF488-Azide (Jena Bioscience) and ATTO 647N maleimide (ATTO-TEC GmbH) were used for fluorescent labeling. Azide was coupled to 5´-alkyne group introduced by Alkylacetylene Pro phosphoramidite (Primetech ALC). An in-house synthesized 4,4´,4\textacutedbl-trimethoxytrityl (TMTr)-protected 5´-Thiol-Modifier C6 phosphoramidite was used for the synthesis of 5´-thiol-DNA oligonucleotides\cite{kupihar_novel_2003}.

DNA oligonucleotides (Table \ref{table:oligomer_synthesis}) were synthesized using a DNA/RNA/LNA H-16 synthesizer (K\&A Laborgeraete, Schaafheim, Germany) by standard $\beta$-cyano\-ethyl phosphoramidite chemistry at a nominal scale of \qty{0.2}{\umol}. Oligonucleotides were purified by HPLC on an RP-C18 column under ion-pairing conditions (using mobile phases containing \qty{0.1}{mM} triethylammonium acetate (TEAA), pH 6.5 buffer). Quality of the oligonucleotides was checked by HPLC and MS analysis.

\begin{table}
\begin{tabular}{ |c|c| }
 \hline
 \parbox[t]{0.35 \linewidth}{5´-Thiol-DNA docking strand} & \parbox[t]{.65 \linewidth}{5´-SH-C6-AATCTGTATCTATATTCATCATA\\GGAAACACCAAAGATGATATTT\\\textbf{TCTTTAAT}} \\
 \hline
 \parbox[t]{.35 \linewidth}{Complementary DNA for double strand formation} & \parbox[t]{.65 \linewidth}{5´-AAATATCATCTTTGGTGTTTCCT\\ATGATGAATATAGATACAGATT} \\
 \hline
 \parbox[t]{.35 \linewidth}{ATTO647N-DNA imager strand} & \parbox[t]{.65 \linewidth}{\textbf{ATTO647N}-SH-linker-\textbf{AATGAAGA}} \\
 \hline
 \parbox[t]{.35 \linewidth}{AF488-DNA imager strand} & \parbox[t]{.65 \linewidth}{\textbf{AF488}-alkyn-linker-\textbf{AATGAAGA}} \\
 \hline
\end{tabular}
\caption{\label{table:oligomer_synthesis} Sequences of the thiol-modified and fluorescently labeled oligonucleotides}
\end{table}

\subsubsection{Sample preparation for the DNA-PAINT measurements}
\label{sec:experiments_sample_preparation}

The cover slip was first cleansed with plasma cleaner under a stream of nitrogen for 1 hour. \qty{5}{\uL} colloid solution of the Au/Ag alloyed nanoparticles was dropped and dried on the treated cover slip. This step was repeated until the desired nanoparticle density was reached, then the cover slip was rinsed with pure water.

The nanoparticle coated cover slip was mounted on the microscope and selected areas were scanned over with intense, focused laser light. For this purpose, the confocal unit of the microscope was utilized, with an objective with an NA of 1.49, and simultaneous laser illumination of \qty{647}{\nm} and \qty{488}{\nm} was applied with nominal powers of \qty{40}{\mW} and \qty{30}{\mW}, respectively. This preparation step resulted in the metallic multi-crystalline nanoparticles being melted into spherical shapes (Supplementary Fig. S2). The alteration of the nanoparticles was checked using white light transmission illumination.

For the functionalization procedure, the protocol of Steuwe was followed\cite{steuwe_visualizing_2015}. Thiol-terminated oligos were dissolved at a concentration of \qty{1}{\micro M} in \qty{10}{mM} Tris buffer at a pH of 7.0, containing a high ionic concentration of \qty{1}{M} NaCl. To cleave the dithiol bonds, \qty{0.01}{M} concentration of TCEP was also added. \qty{60}{\uL} of the solution was dropped onto the nanoparticle coated cover slip, which was covered and left in a refrigerator overnight to minimize evaporation. The next day, the surface was washed again with the aforementioned Tris buffer (containing \qty{1}{M} NaCl at \qty{10}{mM}, pH=7.0). After rinsing, \qty{60}{\uL} drop of the complementary oligomer strand solution (\qty{100}{nM} concentration prepared in \qty{10}{mM}, pH=7.0 Tris buffer containing \qty{1}{M} NaCl) was placed on the cover slip. The drop was left on for 2 hours and was then rinsed off.

For the control samples, MPTES silanized coverslips were prepared using modified protocols of Verdoold\cite{verdoold_scattering_2012} and Haddada\cite{ben_haddada_optimizing_2013}. A cover slip was placed in a heatproof, sealable container containing \qty{50}{mM} MPTES solution prepared in pure methanol. The container was put into the oven at \qty{75}{\degreeCelsius} for \qty{12}{\hour} to achieve the silanization of the glass surface. After the silanization step, the sample was washed twice in anhydrous methanol, dried under air, and further heated at \qty{90}{\degreeCelsius} for \qty{2}{\hour} in air. All further functionalization steps matched those of the nanoparticle coated cover slips.

\subsubsection{Bright field/SEM calibration of the nanoparticle size}
\label{sec:experiments_calibration}

The metallic nanoparticles larger than $\sim\qty{50}{\nm}$ can be detected by simple bright field optical microscopy. When brought into focus, the metallic nanoparticles are seen as diffraction limited dark spots (Figure \ref{fig:methods_experiments}A), whose contrast rapidly increases with the particle size, while dielectric contaminations have low visibility and are brighter than the background. Exploiting this behavior allows one to identify the metallic nanoparticles and discern between the different sized particles. For slightly better visibility, 520/40 bandpass filter was placed in the emission path.

On the bright field images, visibility was quantified as the Weber contrast calculated from the from the fitted Gaussian volume and from the background level of a single camera pixel. However, the contrast of the nanoparticles on the bright field images did not directly provide their sizes. To overcome these challenges, both bright field optical and SEM images were taken on the same areas containing the same laser-treated nanoparticles. By measuring the contrast from the bright field images and the diameters from the SEM images (Figure \ref{fig:methods_experiments}B) of individual nanoparticles, we acquired particle-size-visibility calibration data on which we fitted the calibration curve (Figure \ref{fig:methods_experiments}C). For the SEM images, a Hitachi S-4700 scanning electron microscope was used with \qty{10}{\kV} acceleration voltage and \qty{10}{\uA} current. Before the measurements, the sample was covered with a thin gold layer to eliminate the electrical overload of the samples. The pictures were registered by using 2k, 15k, and 40k magnifications on the same slide.

\subsubsection{Instrumentation and methods used for the DNA-PAINT measurements}

The DNA-PAINT measurements were performed on a custom-made inverted microscope based on a Nikon Eclipse Ti-E frame. The applied laser beams were focused onto the back focal plane of the microscope objective (Nikon CFI Apo 100x, NA=1.49), which produced a collimated beam on the sample. All the SMLM images were captured with a linearly polarized beam and TIR illumination at an excitation wavelength of \qty{488}{\nm} (Cobolt Coherent Sapphire 488 LP) or \qty{647}{\nm} (MPB Communications Inc., $P_{max}=\qty{300}{\mW}$), with \num{0.48} and \qty{0.85}{kW/cm^2} intensities, respectively. The laser intensity was controlled via an acousto-optic tunable filter (AOTF). Images were captured with an Andor iXon3 897 EMCCD camera (512x512 pixels with \qty{16}{\um} pixel size), whose lowest readout noise normal mode was utilized with \qty{100}{\ms} exposure time, thanks to the relatively slow dynamics of DNA-PAINT. From the individual ROIs, 10,000-20,000 camera frames were captured with reduced image size (crop mode). Excitation and emission wavelengths were spectrally separated with a fluorescence filter set (LF\-405/\-488\-/561\-/635-A-000, Semrock Inc.) and using an additional emission filter (FF01-520/44-25 or BLP01-647R-25, Semrock Inc.) in the detector arm. During the measurements, the perfect focus system of the microscope was used to keep the sample in focus with a precision of $<\qty{30}{\nm}$.

Before the DNA-PAINT measurements, we created \qty{1}{nM} solution of imager strands of both dyes in EDTA buffer. We also added \qty{2}{mM} aged Trolox (illuminated for \qty{5}{\minute} with UV light) to reduce the photobleaching and \qty{100}{nM} NaCl to increase the ionic strength. The solution was prepared in $\qty{0.5}{\mL}$ DNA LoBind\textsuperscript{\textregistered{}} PCR tubes (Eppendorf SE) to minimize the adsorption of the oligonucleotids.

The polarization sensitive measurements were performed as described previously\cite{sinko_polarization_2017}. A birefringent wedge placed in the Fourier plane of the image plane separated the two perpendicular linearly polarized components, here termed as the "x" and "y" components. The PSFs of the two components were imaged on the camera separated by a distance corresponding to $\sim\qty{1.6}{\um}$ in the object plane. Beside the displacement of the polarization components in the image plane, no further PSF distortion was introduced by this optical element. Hence a single detector area was used to capture the doubled PSFs, and conventional Gaussian fitting localization algorithms supplemented by localization pairing could be used.

\subsubsection{Data evaluation of the DNA-PAINT measurements}
\label{sec:experiments_evaluation}

The acquired image stacks containing the double PSFs were processed and evaluated with the rainSTORM localization software\cite{rees_elements_2013}. Prior to the localization, the static fluorescent background was removed from the image stacks as described previously\cite{toth_mapping_2020}. This step was necessary for the measurements performed with the $\qty{488}{\nm}$ excitation, as the nanoparticles themselves produced a non-negligible, constant fluorescent background. For the localization procedure, we chose 2D multi-Gaussian fitting with a constant background. To identify the blinking events, we applied increased Gaussian blurring to mitigate the effects of the PSF distortion induced by the nanoparticles (to avoid fitting two Gaussian functions to a single, distorted PSF).

\begin{figure}[h]
\centering
\includegraphics[width=\linewidth]{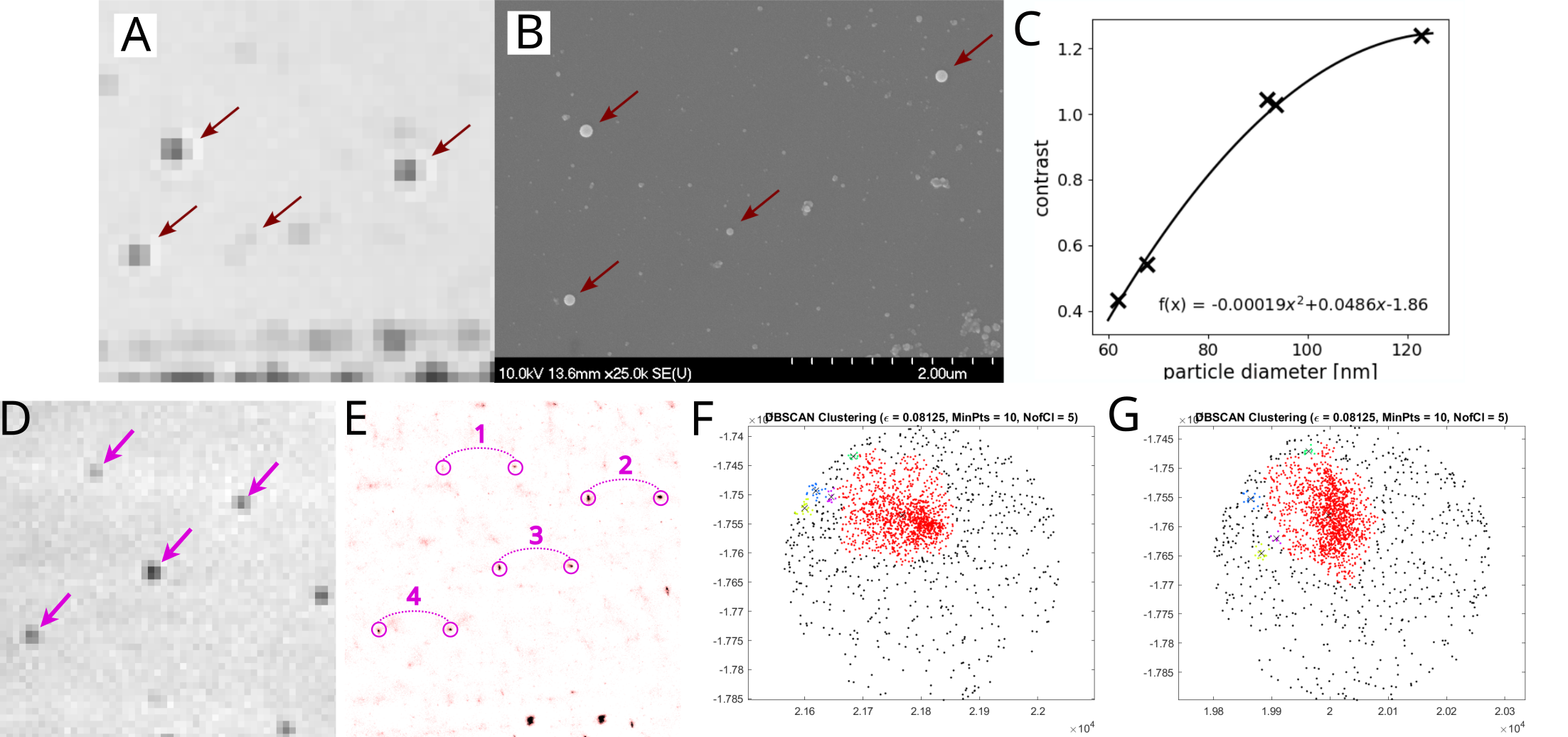}
\caption{A:Laser treated spherical nanoparticles on light microscopy image. B: The same nanoparticles on SEM image. C: Calibration curve based on the measured sizes with SEM and on the measured contrast with light microscopy. D: Functionalized nanoparticles identified with transmission light microscopy. E: Spot pairs corresponding to nanoparticles on the localization image. F-G: Clustered localizations within a spot pair of a nanoparticle.}
\label{fig:methods_experiments}
\end{figure}

Once the localization process was finished, small regions were cut out around the double images of the nanoparticles (Figure \ref{fig:methods_experiments}D, E). The nanoparticles were identified on the bright field images and were selected by the corresponding spots on superresolved localization images. Within these ROIs, DBSCAN\cite{simoudis_international_1996} clustering was performed on the localization data with $\varepsilon = \qty{5}{\nm}$ and $N = 1$ parameters, and the largest clusters were selected from each ROI (Figure \ref{fig:methods_experiments}F, G). The ROI selection and the subsequent clustering ensured that the majority of the selected localizations belonged to the examined nanoparticle, while the blinking events occurring on the glass surface were effectively discarded. All additional localization filtering was disabled, so even the feeble PSFs of the highly polarized blinking events were also taken into account. In the final step, pair finding was performed on the localizations within the accepted clusters, i.e. the localizations belonging to the same blinking events were paired. From the ratio of the camera signals of the localization pairs, the blinking events’ polarization degrees could be calculated. Once we obtained the polarization degrees of the blinking events, we could create the histogram of the polarization degree distributions of the individual nanoparticles.

\section{Results}
\label{sec:results}

\subsection{Calculations}
\label{sec:results_calculations}

\newcommand{\vparallel}{\rotatebox[origin=c]{90}{$_{\parallel}$}}

Our calculations show that the interaction of dyes with the plasmonic nanoparticles induce significant fluorescence polarization to the freely rotating dye’s emission. This statement holds even though the study was performed with spherical particles and with rotationally unconstrained dyes in the slow rotation limiting case, i.e. when the dyes emit in the same dipole orientation as they are excited. The emergence of polarization anisotropy can be explained at emission level. Apart from the dye, the nanoparticle also acts a source of radiation as a results of the oscillating charge distribution induced by the dye’s dipole moment. The detectable signal is produced by the coherent emission of the nanoparticle-dye coupled system\cite{goldwyn_mislocalization_2018,zuo_model-based_2021} (Supplementary Fig. S3). Consequently, the magnitude and phase relation of the two sources and its orientation dependence determine the polarization state of the freely rotating dye’s emission.

\begin{figure}[h]
\centering
\includegraphics[width=\linewidth]{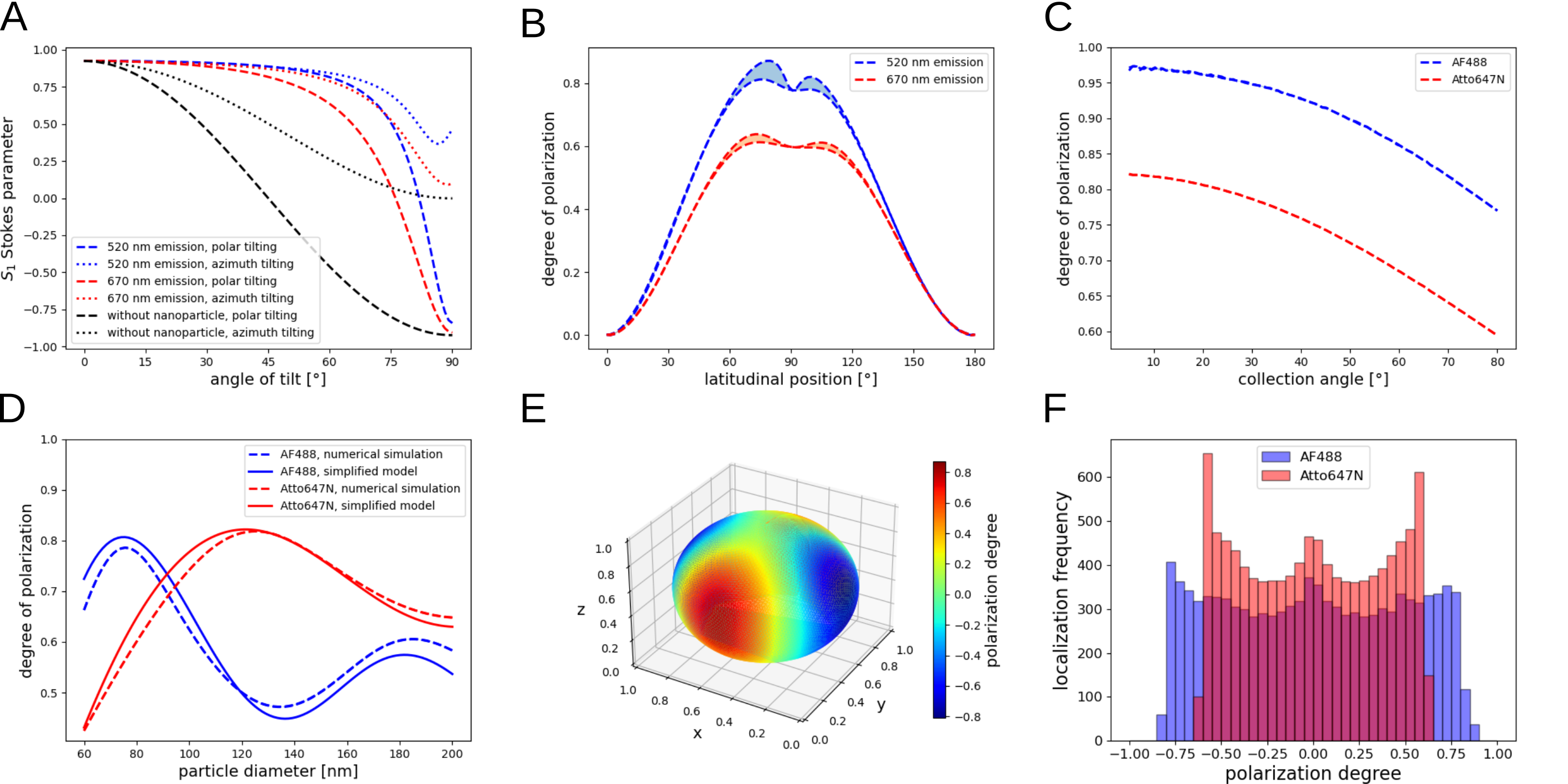}
\caption{A: The (normalized) S\textsubscript{1} Stokes parameter of the emission as a function of polar (dashed line) or azimuth (dotted line) angle tilt. Its absolute value provides the degree of polarization. The dyes are located on the "x" axis. B: The degree of polarization’s dependence on the dye latitudinal position. The area between the lines shows the possible values of dyes in different longitudinal positions. C: The degree of polarization of the collected emission as a function of the collection angle with the AF488 and the Atto647N dyes located in the "x-y" plane. D: The calculated degree of polarization values of dye emissions in the "x-y" plane as a function of the nanoparticle size. The solid line depicts the prediction of the simplified model, the dashed lines shows the simulated degree of polarization values. E: The measurable polarization degree (MPD) distribution of different dye positions on the sphere around the nanoparticle in the case of the AF488 dye. F: Simulated MPD histograms of the AF488 and Atto647N dyes. All calculation (with the exception of the subfigure D) was performed with a \qty{80}{\nm} diameter particle.}
\label{fig:results_theoretical}
\end{figure}

This phenomenon can be interpreted in a simplified view using the induced mirror charges\cite{ropp_nanoscale_2015,bloksma_imaging_2021}. The emitting dye, as an dipole moment, induces an image dipole in the adjacent conductive surface. The direction of the surface perpendicular component of the image dipole matches that of the dye, however the surface parallel component is opposite. So while the perpendicular component of the two sources interfere constructively, the parallel components attenuate each other, promoting the absorption process instead of the emission. Although this model strictly holds only for static dipole near a planar metallic surface, many aspects translate to the dye-nanoparticle system. Out of the two dyes investigated, the aforementioned model only break for the lower wavelength AF488’s emission in case of large particles (above $\sim\qty{120}{\nm}$), when the higher modes also gain significance (Supplementary Fig. S4). Beyond the mirror charge model, the dye induces a dipole moment parallel to the nanoparticle center–dye position line (perpendicular to the particle surface) with higher propensity. Due to these effects, the polarization of the emission no longer corresponds to the dipole orientation\cite{zuo_rotation_2019}, but is skewed towards the nanoparticle center–dye position line (Figure \ref{fig:results_theoretical}A) for most dye orientation. Consequently, even when the dye possesses high rotational mobility and its emission must be averaged in a wide orientation range, the collected emission remains highly polarized. The collected emission’s degree of polarization is highest when the dye is placed around the "x-y" plane and only vanishes near the top and bottom of the particle (Figure \ref{fig:results_theoretical}B), i.e. it is highly sensitive to the relative alignment the dye and the direction of the emission collection. To a lesser extent, the degree of polarization also decreases monotonically with increasing collection angle (Figure \ref{fig:results_theoretical}C).

Here we note that the degree of polarization and the total radiative rate show the same pattern as the particle size changes (Supplementary Fig. S5). This qualitative behavior can be explained with the fact that the more the nanoparticle contributes to the emission, the higher the polarization anisotropy is. Based on this observation, here we propose a simplistic model describing the degree of polarization of the nanoparticle-dye coupled system. In the model, we treat the dye and the nanoparticle system as a single dipole moment, and neglect the vectorial nature of the excitation field (for the effects of the approximation see the Supplementary Fig. S6). The resultant dipole’s components are approximated to be linearly proportional to the dye’s dipole surface perpendicular and parallel components multiplied by the square root of the corresponding radiative decay rates:
$$
\vec{d} = \left({d}_\perp , {d}_{\parallel} , {d}_{\vparallel}\right) \sim
(
\sqrt{{\sigma}_{\perp}^ {rad}} \cos{\theta },
\sqrt {{\sigma}_{\parallel}  ^{rad}} \sin{\theta} \cos{\phi},
\sqrt {{\sigma}_{\parallel}^{rad}} \sin{\theta} \sin{\phi}) \mathrm{,}
$$
where $\vec{d}$ is the resultant dipole vector, ${d}_\perp$, ${d}_{\parallel}$, ${d}_{\vparallel}$ are its radial, latitudinal and longitudinal components, respectively, $\theta$ and $\phi$ are the dipole vector’s polar and azimuth angles (Figure \ref{fig:methods_theoretical}B), ${\sigma}_{\perp}^ {rad}$ and ${\sigma}_{\parallel}^{rad}$ are the radiative rates of surface perpendicular and parallel dipoles, respectively, for whose calculation there are well known formulas\cite{kerker_surface_1980,ruppin_decay_1982,chew_transition_1987,mertens_plasmon-enhanced_2007}.

In our model, these dipole components separately contribute to the far field emission’s polarization components, in other words, their interference integrated over the emission collection is neglected. By exploiting the system’s symmetry and by choosing a proper coordinate system one only has to determine the S\textsubscript{1} Stokes parameter of the polarization state to calculate the degree of polarization:

$$
{I}_{x} =
{R}_{\perp}^{x} {p}_{\perp} {CE}_{\perp} +
{R}_{\vparallel}^{x} {p}_{\vparallel} {CE}_{\vparallel} +
{R}_{\parallel}^{x} {p}_{\parallel} {CE}_{\parallel}
$$

$$
{I}_{y} =
{R}_{\perp}^{y} {p}_{\perp} {CE}_{\perp} +
{R}_{\vparallel}^{y} {p}_{\vparallel} {CE}_{\vparallel} +
{R}_{\parallel}^{y} {p}_{\parallel} {CE}_{\parallel}
$$

$$
P = \frac{|{I}_{x}-{I}_{y}|}{{I}_{x}+{I}_{y}} = |S_1| \mathrm{,}
$$
where $p$ describes the chances of photon emission by the given dipole component, $R_x$ and $R_y$ is the ratio of the collected "x" and "y" linearly polarized emission, and $CE_{\perp,\parallel,\vparallel}$ are the collection efficiencies of the given dipole mode, $P$ is the degree of polarization. The $R$ and $CE$ values can be determined from numerical calculation using free dipole emitters (without any nanoparticle). In our case ${R}_{\perp}^{x} = {R}_{\vparallel}^{y} = 0.96$, ${R}_{\vparallel}^{x} = {R}_{\perp}^{y} = 0.04$, ${R}_{\parallel}^{x} , {R}_{\parallel}^{y} = 0.50$ and ${CE}_{\perp} = {CE}_{\vparallel} = {{CE}_{\parallel}}/{0.86}$. In the slow rotation limit, the dipole component’s photon emission chances can be calculated as:
$$
p_{\perp} =
\int_{0}^{2\pi} \int_{0}^{\pi}
\frac
{\sigma_{\perp}^{rad}\cos^2{\theta}}
{\sigma_{\perp}^{tot}\cos^2{\theta}+\sigma_{\parallel}^{tot}\sin^2{\theta}+\frac{1}{Q_0}-1}
\frac{\sin{\theta}}{4\pi}
\,d\theta\,d\phi
\mathrm{,}
$$
$$
p_{\vparallel} =
\int_{0}^{2\pi} \int_{0}^{\pi}
\frac
{\sigma_{\parallel}^{rad}\sin^2{\theta}\sin^2{\phi}}
{\sigma_{\perp}^{tot}\cos^2{\theta}+\sigma_{\parallel}^{tot}\sin^2{\theta}+\frac{1}{Q_0}-1}
\frac{\sin{\theta}}{4\pi}
\,d\theta\,d\phi
\mathrm{,}
$$
$$
p_{\parallel} =
\int_{0}^{2\pi} \int_{0}^{\pi}
\frac
{\sigma_{\parallel}^{rad}\sin^2{\theta}\cos^2{\phi}}
{\sigma_{\perp}^{tot}\cos^2{\theta}+\sigma_{\parallel}^{tot}\sin^2{\theta}+\frac{1}{Q_0}-1}
\frac{\sin{\theta}}{4\pi}
\,d\theta\,d\phi
\mathrm{,}
$$
where $\sigma_{\perp}^{tot}$ and $\sigma_{\parallel}^{tot}$ denote the total rates (including both the absorption and emission processes) of surface perpendicular and parallel dipoles, respectively. See the Supplementary Eq. 8 and 9 for the evaluated formulas. The model agrees well with the degree of polarization values of the AF488 and Atto647N dyes served by the numerical calculations and predicts well their tendencies as a function the nanoparticle size (Figure \ref{fig:results_theoretical}D). The predicted degree of polarization is slightly overestimated with the exception of the larger ($>\qty{120}{\nm}$ diameter) nanoparticles in case of the AF488 dye.

Polarization sensitive measurements are typically performed by splitting the emission into two polarization channels with a single polarizing element. This method is not suitable for the acquisition of the full polarization state of the emission, only the S\textsubscript{0} and S\textsubscript{1} Stokes parameters can be obtained. For symmetry reasons, the emission near the spherical nanoparticles cannot be circularly polarized (S\textsubscript{3}), it still leaves one Stokes parameter (S\textsubscript{2}) hidden. For this reason, the MPD depends on the dye’s binding longitudinal position on the particle relative to the polarizing element’s axes as well as on its latitudinal position (Figure \ref{fig:results_theoretical}E). In order to draw conclusions from the MPD values, Monte Carlo simulations were performed and MPD histograms were created (Figure \ref{fig:results_theoretical}F). The maximum and minimum values of the MPDs were determined by the emission’s degree of polarization. All simulated histograms show a sharp cut-off at these boundary values. Between them, the distribution is relatively flat, however, the shapes has wavelength and nanoparticle size dependence. We note that the polarization distribution histogram possess near reflection symmetry around zero, which is slightly broken by the excitation’s contribution.

\subsection{Experiments}
\label{sec:results_experiments}

In order to make deductions about the underlying interactions, we measured the polarization degrees of single-molecule emissions on nanoparticles with both the AF488 and Atto647N dyes. However, the strength of the plasmonic interaction cannot be directly inferred from the measured polarization degree of the individual blinking events, because other parameters, like the analyzer alignment or the dye binding latitudinal position on the spherical nanoparticle also play a major role in the measured value. We can gain the sought after information only by recording a series of blinking events, and comparing the resultant polarization degree distribution with elaborate calculations.

\begin{figure}[h]
\centering
\includegraphics[width=\linewidth]{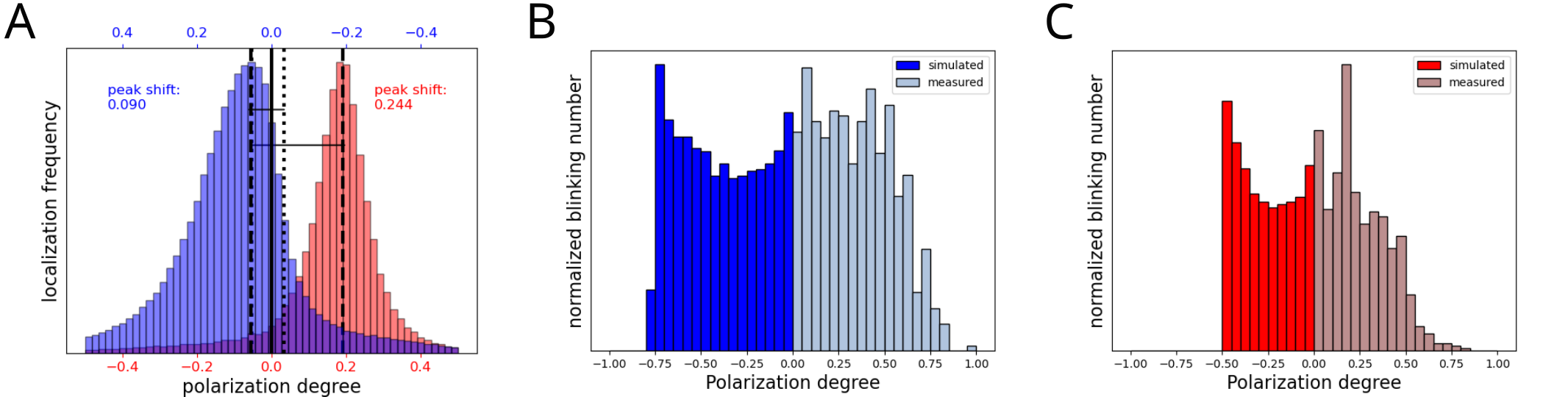}
\caption{A: Polarization degree distribution of the control measurement of the AF488 (blue) Atto647N (red) dyes with s-polarized TIR excitation. B and C: Measured and simulated polarization degree distribution of the AF488 and Atto647N dyes on a $\sim\qty{55}{\nm}$ diameter nanoparticle, respectively. Note that, while the control measurements are depicted in the
$[-0.5, 0.5]$ range, the histograms of the nanoparticles are shown in the $[-1.0, 1.0]$ polarization degree range.}
\label{fig:results_experiments}
\end{figure}

The control measurements with p-polarized TIR illumination all served relatively narrow single peaked polarization distributions centered around zero (Supplementary Fig. S7). This indicates that the rotation of the AF488 and Atto647N dyes are highly unconstrained, i.e. their orientations are not limited significantly by the DNA strands they are attached to. We attribute the non-finite width of the polarization degree histograms to the inherent uncertainty of the measurements. In addition, the control measurements were also performed with s-polarized TIR illumination (Figure \ref{fig:results_experiments}A). By introducing an excitation polarization component parallel to one of the axes of the analyzer, a non-zero polarization degree of emission can be measured. From the extent of the shift, we can infer characteristic time ratios of the fluorescence lifetime and rotational correlation time\cite{backer_enhanced_2016}. Assuming unconstrained rotation, we matched the calculated polarization degrees to the experimental values a used modified code of TestSTORM\cite{novak_teststorm_2017}. The results indicate that, without plasmonic interaction, the ratios of the fluorescence lifetime and the rotation correlation time are $3.34$ for the AF488 and $0.59$ for the Atto647N dyes. We also note that the polarization distribution of the AF488 dye is somewhat wider than that of the Atto647N and has slowly decaying tails. This discrepancy can be attributed to the fact that during the \qty{488}{\nm} excitation, beside the AF488 dye, other, dimmer fluorescent molecules could also be observed on the glass surface with relatively high density. Without thresholding the brightness of the localizations, these could not be effectively filtered out, and interfered with the pair finding algorithm.

Polarization sensitive measurement on Au-Ag alloyed nanoparticles were also performed using the AF488 and Atto647N dyes (Figure \ref{fig:results_experiments}B, C). Compared to the control measurements their polarization degree histograms are significantly wider and has flatter distribution instead of pointy peak-like characteristics. This broadening indicates the presence of the plasmonic interaction of the dyes bound to the particles. For the $\sim\qty{55}{\nm}$ diameter particle evaluated, even high polarization degrees of $\sim 0.7$ and $\sim 0.5$ were common for the AF488 and Atto647N dyes, respectively. We also note that, the measured polarization degree histograms does not have as sharp a cut-off at the expected MPD extrema as the calculated ones. We attribute this to the uncertainties of the measurements, and it hinders the extraction of well-defined extreme values from the measurements. Nonetheless, the width of the measured polarization degree histograms of the two dyes follow the tendency predicted by the calculations.

These results shows that the relatively moderate interaction with less than $\sim10\times$ radiative rate enhancement can result in relatively high measurable polarization degree values. In conclusion, the polarization degree of the fluorescent dye’s emission is highly sensitive to the presence of plasmonic interaction with the nanoparticles and can be used to infer the interaction strength.

\section{Conclusion}
\label{sec:conclusion}

With elaborate modeling, we described the polarization state of the emission originating from a freely rotating fluorescent dye-plasmonic nanosphere coupled system. Our calculations suggest that even in case of relatively weak plasmonic interaction, we can expect a high degree of polarization whose extent correlates with the strength of the interaction. The origin of this fluorescence anisotropy is that for most dipole orientations the polarization of the collected emission is highly skewed towards the nanoparticle center-fluorophone axis and possesses a higher degree than it would be expected from an uncoupled dipole moment. Furthermore, the degree of polarization of the collected emission is highly dependent on the latitudinal position of the fluorophore. Thus, instead of well-defined values, we can only measure a broad distribution of polarization degrees of single molecule events. We also showed that instead of complicated numerical calculations incorporating dye orientation averaging, it is possible to predict the polarization degree of emission with a simplistic model. The model satisfactorily predicts the degree of polarization values provided by the more complicated numerical calculations in a large nanoparticle size range.

We performed polarization sensitive DNA-PAINT measurements on Au/Ag alloyed plasmonic nanospheres and on control samples without nanoparticles. Although we only measured two perpendicular linear polarization states, which is insufficient to reconstruct the full state of polarization, the resultant polarization degree distributions provided valuable information. The measurements on the nanoparticles resulted in significantly broader and flattened distributions compared to the control measurements, indicating the presence of plasmonic interaction. Furthermore, the distributions of the AF488 and Atto647N dyes followed the tendency predicted by the calculations.

\section*{Author Contribution}

\textbf{Tibor Novák}: Conceptualization, Methodology, Formal Analysis, Investigation, Software, Visualization, Writing - Original Draft, Funding acquisition \textbf{Péter Bíró}: Methodology, Investigation, Data Curation, Formal Analysis, Writing - Review \& Editing \textbf{Györgyi Ferenc}: Methodology, Resources \textbf{Ditta Ungor}: Resources, Writing - Review \& Editing \textbf{Elvira Czvik}: Methodology, Resources \textbf{Ágota Deák}: Investigation \textbf{László Janovák}: Investigation, Funding acquisition \textbf{Miklós Erdélyi}: Supervision, Writing - Review \& Editing, Funding acquisition

\section*{Acknowledgement}

The project (TKP2021-NVA-19) has been implemented with support provided by the Ministry of Innovation and Technology of Hungary from the National Research, Development and Innovation Fund, financed under the TKP2021-NVA funding scheme. Ágota Deák is very thankful for the National Research, Development and Innovation Office-NKFIH (Hungary) for support through program PD 142293. This paper was also supported by the ÚNKP-19-3 -SZTE-212 (T.N.), the ÚNKP-20-4 -SZTE-591 (T.N.), the ÚNKP-23-5 (L.J.) and ÚNKP-23-4 -SZTE-641 (Á.D.) New National Excellence Program of the Ministry for Innovation and Technology from the National Research, Development and Innovation Fund and by the János Bolyai Research Scholarship of the Hungarian Academy of Sciences. Open access funding provided by University of Szeged.

\bibliographystyle{elsarticle-num}
\bibliography{NovakT_polarization_sensitive_SMLM.bib}





\end{document}